\documentclass[useAMS,usenatbib]{mn2e}

\usepackage{float} 
\usepackage{epsfig}
\usepackage{graphicx}
\usepackage[latin1]{inputenc}
\usepackage{latexsym}
\usepackage{amsmath}
\def\simlt{\ \raise -2.truept\hbox{\rlap{\hbox{$\sim$}}\raise5.truept   %
\hbox{$<$}\ }}
\def\simgt{\ \raise -2.truept\hbox{\rlap{\hbox{$\sim$}}\raise5.truept   %
\hbox{$>$}\ }}                                                          %

\def\be{\begin{equation}}
\def\ee{\end{equation}}
\def\newline{\hfil\break}

\def\la{\mathrel{\hbox{\rlap{\hbox{\lower4pt\hbox{$\sim$}}}\hbox{$<$}}}}
\def\ga{\mathrel{\hbox{\rlap{\hbox{\lower4pt\hbox{$\sim$}}}\hbox{$>$}}}}

\def\MS7{MS 0735.6+7421}

\title[Radio halo in the Coma cluster at 4.8 GHz]{High significance detection at 4.8 GHz of the radio halo in the Coma galaxy cluster with the Sardinia Radio Telescope}
\author[P. Marchegiani et al.]{P. Marchegiani$^{1}$\thanks{E-mail: paolo.marchegiani@inaf.it}, M. Murgia$^{1}$, F. Loi$^{1}$, V. Vacca$^{1}$, F. Govoni$^{1}$, S. Cocchi$^{2,1,3}$,\newauthor and F. Gandossi$^{2,1,3}$\\ \\
$^{1}$INAF-Osservatorio Astronomico di Cagliari, Via della Scienza 5, I-09047 Selargius (CA), Italy\\
$^{2}$Dipartimento di Fisica e Astronomia, Universit\`a di Bologna, via P. Gobetti 93/2, 40129 Bologna, Italy\\
$^{3}$INAF, Istituto di Radio Astronomia, Via Gobetti 101, 40129 Bologna, Italy\\
}

\begin{document}

\date{Accepted 2026 February 16. Received 2026 January 12; in original form 2025 October 28}

\pagerange{\pageref{firstpage}--\pageref{lastpage}} \pubyear{2026}

\maketitle

\label{firstpage}

\begin{abstract}
We present the results of observations of the radio halo in the Coma galaxy cluster at 4.8 GHz performed with the Sardinia Radio Telescope. The radio halo in this cluster is detected for the first time at this frequency with a statistical significance higher than $3\sigma$. After the removal of the Radio Frequency Interference and of the discrete sources contribution, and after the correction for the Sunyaev-Zel'dovich effect, we estimate a flux density of $61\pm11$ mJy, higher than the value previously reported in literature at this frequency. By using the value we obtained, it is possible to estimate an integrated spectral index between 4.8 and 6.6 GHz of $\alpha\sim1.17$, where $F(\nu)\propto \nu^{-\alpha}$, indicating a possible higher-frequency slowdown of the spectral steepening observed between 1.4 and 4.8 GHz. Such a spectral behavior is compatible with turbulent re-acceleration if the seed electrons have a spectrum extending up to high energies, as in the case of continuous injection by hadronic interactions or dark matter annihilation. We also report the detection at 4.8 GHz of a polarized spot inside the halo, without an evident counterpart, already detected at 6.6 GHz.
\end{abstract}

\begin{keywords}
galaxies: clusters: individual: Coma - galaxies: clusters: intracluster medium - radio continuum: general - acceleration of particles
\end{keywords}


\section{Introduction}

Radio halos are diffuse emissions located at center of galaxy clusters, usually having angular extension and morphology comparable with those of the hot gas in the Intra-Cluster Medium (ICM; e.g. Feretti et al. 2012). While at the moment radio halos have been found in hundreds of clusters (e.g. van Weeren et al. 2019), the radio halo in the Coma cluster, the first to be discovered (Willson 1970), remains one of the most studied. It is often used as prototype because of its closeness and brightness (e.g. Schlickeiser, Sievers \& Thiemann 1987, Giovannini et al. 1993, Deiss et al. 1997, Thierbach, Klein \& Wielebinski 2003, Brown \& Rudnick 2011, Bonafede et al. 2022, Murgia et al. 2024), and constitutes therefore a target for theoretical studies aimed at determining the properties of relativistic particles and magnetic fields in the ICM (e.g. Jaffe 1977, Rephaeli 1979, En\ss lin \& Biermann 1998, Blasi \& Colafrancesco 1999, Brunetti et al. 2001, Colafrancesco, Profumo \& Ullio 2006, Marchegiani, Perola \& Colafrancesco 2007, Brunetti et al. 2012, Marchegiani \& Colafrancesco 2016, Marchegiani 2019, Nishiwaki \& Asano 2025, Marchegiani et al. 2025).

The flux density of the radio halo in the Coma cluster has been measured at many frequencies (see Murgia et al. 2024 for a recent compilation), allowing to derive a spectrum extending from 30 MHz up to 6.6 GHz. The spectral shape is an important source of information, from which it is possible to obtain constraints about the physics of the cluster, like the origin of the relativistic electrons, the processes of particle acceleration, and possibly the properties of other components of the ICM, like non-thermal protons and dark matter.

An interesting property of the radio halo spectrum in the Coma cluster is that, while it seems to have a quite precise power-law shape up to the frequency of 1.4 GHz (Deiss et al. 1997), after this frequency it seems to have a quick steepening (Thierbach et al. 2003). This steepening is expected because of the effect of radiative energy losses involving high-energy electrons, if the electrons are not continuously produced  
in hadronic interactions between non-thermal protons and thermal nuclei (e.g. Blasi \& Colafrancesco 1999). 
Therefore, the presence of this steepening is interpreted as an evidence against a secondary origin for non-thermal electrons, 
and instead favors other processes,  
as the re-acceleration related to the cluster turbulence that can develop following a major merger (e.g. Brunetti et al. 2001).

Recently, the Coma radio halo has been observed by Murgia et al. (2024) at a frequency never reached in the past, 6.6 GHz, with the Sardinia Radio Telescope (SRT; Prandoni et al. 2017). Interestingly, this observation has shown that the quick spectral steepening observed between 1.4 and 4.85 GHz seems to slow down after 4.85 GHz.
In fact, the spectral index calculated, with the definition $F(\nu)\propto \nu^{-\alpha}$, using the flux density values in table 3 of Murgia et al. (2024) between 1.49 and 4.85 GHz is $\alpha_{1.49}^{4.85}=2.19\pm0.32$, and between 4.85 and 6.6 GHz is $\alpha_{4.85}^{6.6}=-0.9\pm1.4$ (i.e. an inverted spectrum but with an error bar compatible with a flat spectrum).
 This behavior is potentially very significant, because it might be due to the fact that at these frequencies we are observing electrons with energies for which the lifetime due to the energy losses is shorter than the re-acceleration characteristic time (e.g. Brunetti \& Lazarian 2011), possibly providing information on the properties of the seed electrons (Marchegiani et al. 2025).

However, characterizing the properties of radio halos in this range of frequencies is a hard task. In fact, they usually have a steep spectrum, and therefore at high frequencies their surface brightness is weak. In particular, the radio halo in the Coma cluster, which is a very rich and nearby cluster,  
has an angular extension larger than 1 degree (Bonafede et al. 2022). Therefore at high frequencies it is not suitable to be studied with interferometers because of the lack of sensitivity on large scales due to their non-zero minimum baseline. On the other hand, the confusion noise affecting single dish instruments makes it challenging to detect the halo surface brightness, especially in the peripheral regions of the cluster where it is expected to be weaker than at the center. For this reason, single dish observations with large diameter telescopes, and with observing times appropriate to reach a noise level close to the expected confusion noise, are necessary.

At the moment, the only observation of the Coma radio halo around 4.85 GHz available in literature has been presented by Thierbach et al. (2003), who studied the radio halo with the 100 m Effelsberg telescope. Their observations confirmed the difficulty of detecting the radio halo at this frequency, because they did not obtain any detection of the surface brightness of the diffuse emission over the $3\sigma$ level, but just found regions with lower significance surface brightness,  
estimating from these regions a flux density of $26\pm12$ mJy. Moreover, the presence of two bright radio sources close to the cluster center, and of other sources whose spectrum at high frequencies is not well constrained, has made it difficult to separate the emission of the radio halo from the discrete sources (see discussion in Thierbach et al. 2003). Furthermore, at this frequency the Sunyaev-Zel'dovich effect (SZE) can provide a non-negligible contribution, making it necessary to correct the flux density for this effect (e.g. En\ss lin 2002, Brunetti et al. 2013, Murgia et al. 2024). This correction is strongly dependent on the size of the region where the halo is detected, and requires a precise subtraction of the image baseline level, which is complicated in the case of a strong contamination from Radio Frequency Interference (RFI) or if the field of view is small.

In this paper we present the results of observations of the radio halo in the Coma cluster performed using the receiver in the C-low band at SRT, which can access a band centered at 4.9 GHz with a bandwidth of 1.4 GHz (Orfei et al. 2024; Schirru et al. 2026).  
These observations are motivated by two facts. First, as mentioned, the results available in literature at this frequency (Thierbach et al. 2003) are derived with maps where the diffuse emission is not detected at a level over $3\sigma$. The reported rms of those maps, 0.67 mJy/beam, is higher than the confusion noise of 0.3--0.4 mJy/beam 
expected at this frequency  
according to the analytical approximation of Condon (2002) or the numerical simulations of Loi et al. (2019)
for a beam of 2.45 arcmin, as in the case of the Effelsberg telescope. 
Therefore our goal is to check if, by reaching an rms comparable with the confusion noise by adopting an adequate observing time, even with a smaller telescope like the 64 m SRT it may be possible to obtain a more solid detection of the diffuse emission. Second, a direct comparison between the data at 6.6 GHz (Murgia et al. 2024) with those at 4.8 GHz would be more reliable if performed by adopting methods for measuring the flux density and for subtracting the discrete sources contribution that are as similar as possible at the two frequencies, and this can be done more easily by using images that are in our full control. 

Another interesting result presented in Murgia et al. (2024) was the detection at 6.6 GHz of a polarized spot inside the halo, without an evident counterpart in other spectral bands. Since these kinds of polarized structures are expected to be sometimes observable inside the radio halos, with increasing fractional polarization at high frequencies (Govoni et al. 2013), it is interesting to observe the polarization properties of the radio halo in Coma at 4.8 GHz in order to check the possible presence of this spot also at this frequency.

The paper is structured as follows: we present in Sect.2 the details and the results of the observations, and the procedure followed to obtain the image of the radio halo by mitigating the effect of the RFI. In Sect.3 we estimate the flux density of the radio halo, by removing the contribution of the discrete sources and correcting for the SZE. In Sect.4 we discuss the impact of the result we obtain on the high-frequency spectral shape of the radio halo, providing a basic outline of the possible consequences on the information we can derive on the origin of the electrons producing the halo. In Sect.5 we present the polarized intensity image of the radio halo, discussing the properties of the polarized spot. Finally, we summarize our results in Sect. 6. In the paper we adopt a cosmological model with $\Omega_m = 0.308$, $\Omega_{\Lambda} = 0.692$, and $H_0 =67.8$ km s$^{-1}$ Mpc$^{-1}$ (Ade et al. 2016); for this cosmological model the luminosity distance of the Coma cluster, which is located at $z=0.023$, is $D_L=104$ Mpc, and 1 arcmin corresponds to 28.9 kpc.

\section{Observations and data analysis}

We observed the region of the Coma galaxy cluster with the C-low receiver at SRT following an accepted proposal (Program 10-25, P.I. Marchegiani) for a total of 32 hours of observing time (including setups and calibrators) presented at the call for the second semester of 2025; details of the observations are in Table \ref{tab.obs}.

\begin{table*}{}
\caption{Details of the SRT observations}
\begin{center}
\begin{tabular}{|*{8}{c|}}
\hline 
Frequency & Spectral    & Total time  & Center of pointing &  OTF scans & FOV & Program & Observation dates \\
 (GHz) & resolution &  on source & (J2000) & & & &\\
\hline 
4.2--5.4 & 1.46 MHz & 27h & 12:59:23 +27:54:38 & 10RA$\times$10DEC & 2.25$^\circ\times2.25^\circ$ & 10-25 & 8, 9, 10, 15 July 2025 \\
\hline
 \end{tabular}
 \end{center} 
 \label{tab.obs}
 \end{table*}
 
Data have been acquired in four days of observation, with sessions of eight hours each. We used the SArdinia Roach2-based Digital Architecture for Radio Astronomy (SARDARA) backend at SRT (Melis et al. 2018) with 1024 channels in full-Stokes mode, with an effective bandwidth of 1.2 GHz and spectral resolution of 1.46 MHz, centered at the frequency of 4.8 GHz. Apart from the on-source time, we observed 3C286 as primary flux and polarization calibrator, and 3C84 for setting the pointing and the focusing of the telescope, and as polarization leakage calibrator.

We performed on-the-fly map scans of the Coma cluster region, scanning a square region centered on the cluster center and having a side of 2.25 degrees. This size has been chosen in order to reach regions of the sky that are not strongly contaminated by the SZE due to the cluster, allowing for a better subtraction of the baseline level and a better correction for the SZE itself (see Sect.3). We chose the separation between two subscans so that the expected size of the beam (3.9 arcmin) was sampled with four subscans. By sampling at the speed of 6 arcmin/sec, the total time for completing a map was around 80 minutes. In this way, we could complete five maps per session, for a total of twenty maps, equally divided in maps along RA and along DEC.

The data have been reduced using the SCUBE software (Murgia et al. 2016), following a standard procedure involving flagging of bad channels, baseline removal in the calibrator data, calculation of the bandpass correction, imaging of the calibrator, and derivation of the conversion factor between counts and mJy/beam units using the Perley \& Butler (2017) flux scale. 

The data was heavily contaminated by RFI; in order to mitigate this effect, we proceeded in the following manner. Initially, we adopted an automatic procedure for flagging channels with anomalous emission compared to other channels and to other scans. After applying these flags, we calculated the baseline level by fitting it with a linear polynomial, and removed it. We then gridded the maps and combined different scans obtaining a first image, which we inspected carefully for manually flagging the channels with evident contamination from RFI. We applied these flags and recalculated the baseline level, by masking the regions of the radio halo and the discrete sources visible in the image, so that they would not contaminate the fitting of the baseline, using this time a quadratic polynomial. After removing the baseline as estimated in this way, we obtained a new image, which again we inspected to identify other bad channels and bad scans to remove. Finally, we used this image as a model to proceed to a new flagging, identifying channels with evident departures from the model. After this flagging, we estimated and subtracted again the baseline level, and gridded the maps and combined the scans to obtain the final maps in total intensity and polarization  
using weighted stacking (Murgia \& Fatigoni 2024). 
In this procedure, the images in total intensity have been obtained from the circular polarizations $R$ and $L$ as $I=R+L$, and the images in polarization have been derived from Stokes parameters $U$ and $Q$ as $P=\sqrt{U^2+Q^2}$ for polarized intensity and $\Psi=0.5 \arctan (U/Q)$ for polarization angle. 

The maps resulted to have a FWHM beam, as measured by fitting some point sources in the field, of 3.9 arcmin, as expected at this frequency for the SRT diameter.
The noise was estimated by analyzing the rms fluctuations in apparently empty regions of the map, obtaining $\sigma_I\sim1$ mJy/beam in total intensity, of the order of the confusion noise that can be estimated at this frequency for this beam (Condon 2002, Loi et al. 2019), and $\sigma_P\sim0.5$ mJy/beam in polarization. In the following, we present the results in total intensity, while the polarization map is presented in Sect.5.

We checked the consistency and the goodness of the described procedure by measuring the flux densities of some of the sources in the field, external to the halo region to avoid contaminations, in order to compare our results with  
the Very Large Array (VLA) results presented by Bonafede et al. (2010). 
We estimated flux densities by fitting the sources 
with two-dimensional Gaussian functions. We obtained for the source 5C4.42 a flux density of $65.2\pm1.1$ mJy, while the VLA result at 4.9 GHz was 57.8 mJy. For the source 5C4.114 we obtained $16.1\pm1.0$ mJy, with the VLA result being 14.9 mJy. For the source 5C4.127 we obtained $70.5\pm1.5$ mJy, while the VLA value was 72.9 mJy. Therefore the differences are smaller than 10\%, and it seems 
there is no systematic trend in them. 
This suggests they are not due to systematic errors in the calibration, but 
might arise due to fluctuations in RFI or intrinsic variability of the sources. 

In order to have an estimate of possible systematic errors on the flux density  
expected for our observing time, we also estimated  
the flux density of the source 5C4.42 by using the maps separately derived for each of the days of observation. 
We derived four values having a mean value of 61 mJy and a standard deviation of 13 mJy, providing a standard error on the mean of $\sim 13/\sqrt{4}=6.5$ mJy, i.e. a relative error of $\sim10\%$. 
We assume this value as an estimate of systematic errors for our observing time.

 \section{Properties of the radio halo}
 
\begin{figure*}
\centering
\begin{tabular}{c}
\includegraphics[width=0.5\textwidth, trim={1.5cm 9.5cm 0cm 0cm}, clip]{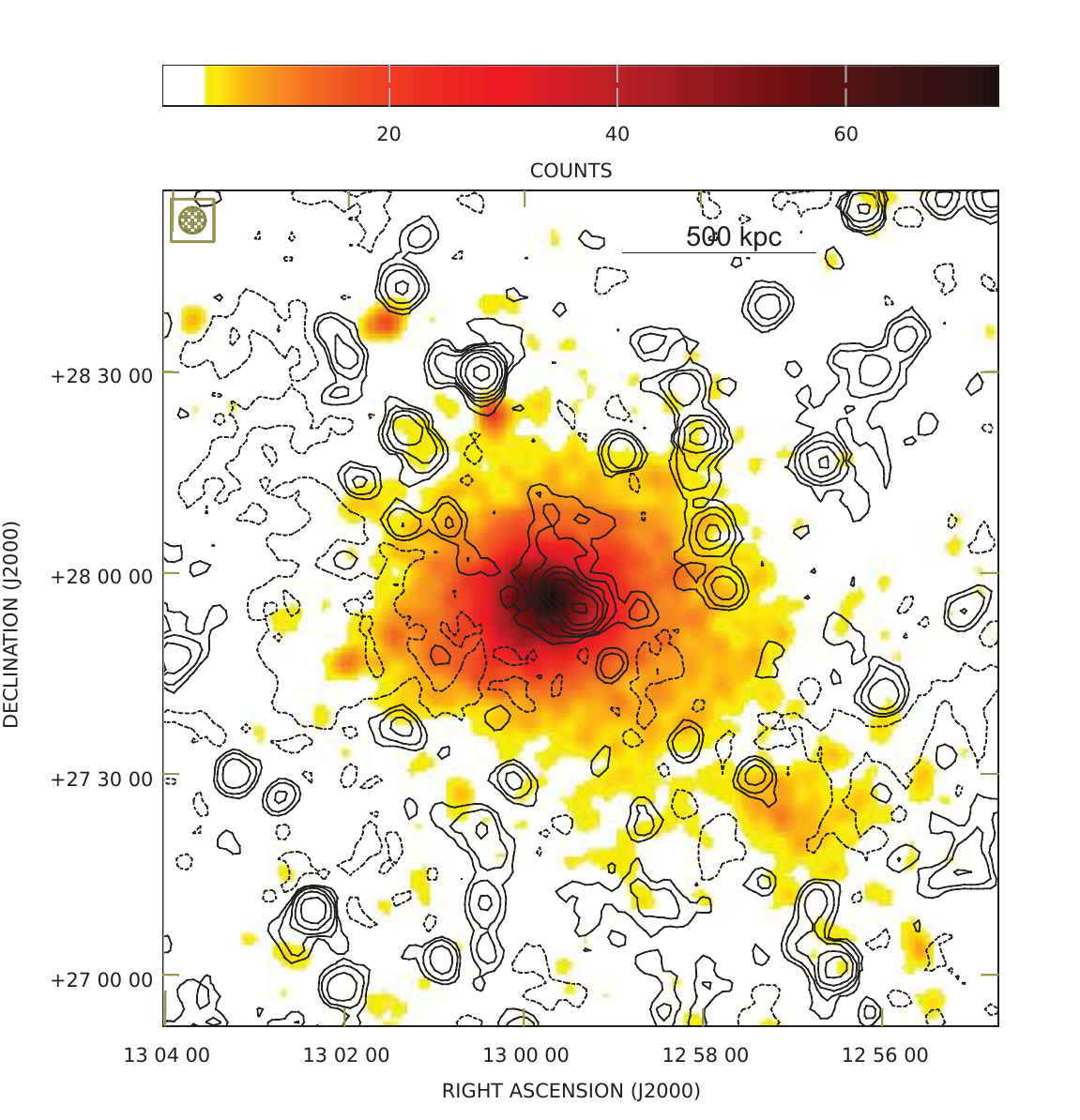}
\includegraphics[width=0.5\textwidth, trim={1.5cm 9.5cm 0cm 0cm}, clip]{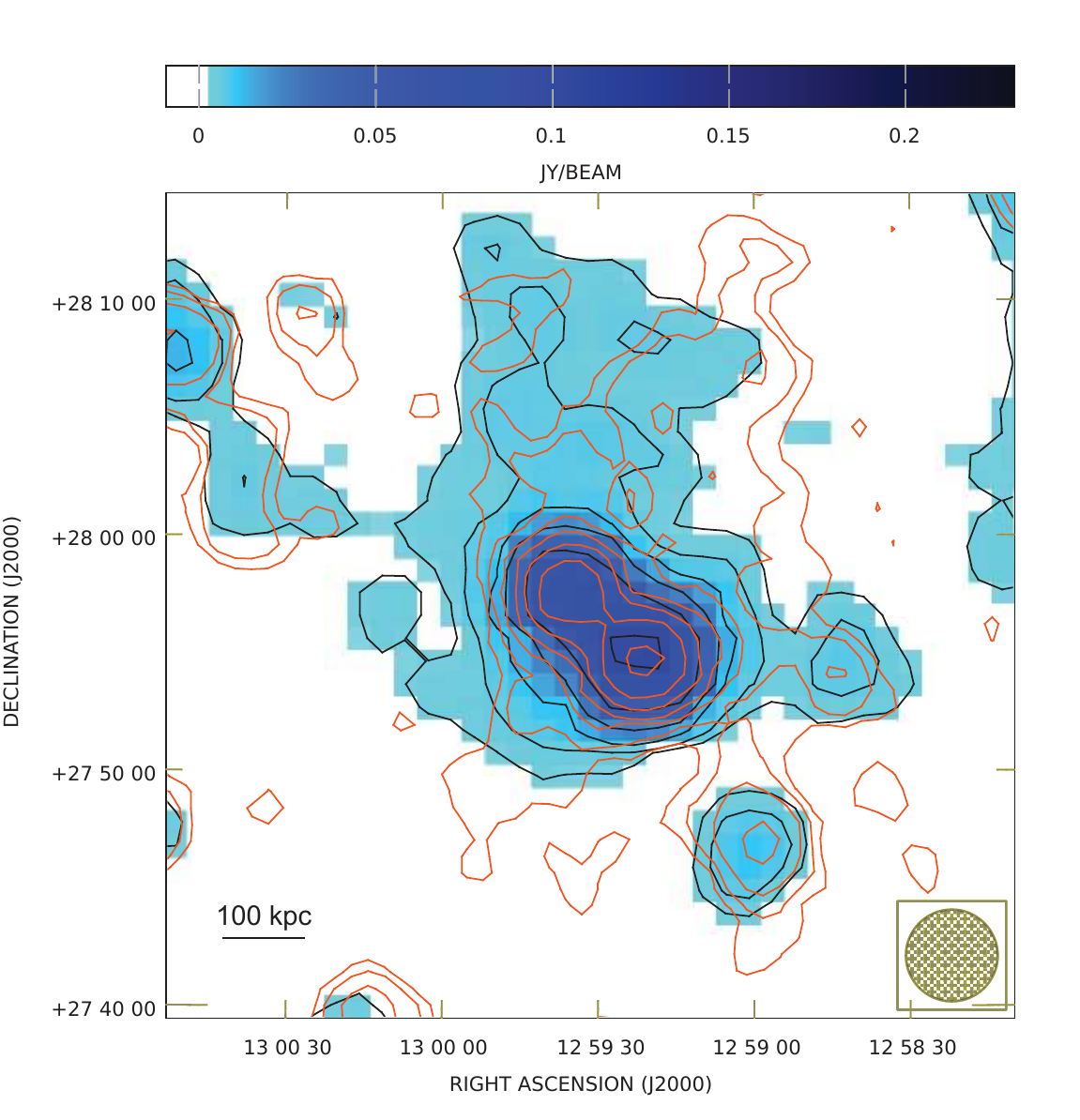}
\end{tabular}
\caption{\textit{Left panel:} 0.1--2.4 keV RASS image, smoothed with a 5-pixel (corresponding to 225 arcsec) Gaussian beam, with SRT contours at 4.8 GHz overlapped. The horizontal bar indicates a length of 500 kpc at the cluster distance. SRT contours are traced at levels of $-3\sigma$ (dotted lines), $3\sigma$, and scale with a factor of two (solid lines), with $\sigma=1$ mJy/beam. The SRT beam (FWHM 3.9 arcmin) is plotted in the top-left corner. \textit{Right panel:} SRT image at 4.8 GHz of the cluster radio halo region with overlapped contours at 4.8 GHz (black lines) and 6.6 GHz (orange lines, from Murgia et al. 2024). Contour levels start at $3\sigma$, with $\sigma=1$ mJy/beam at 4.8 GHz and $\sigma=0.33$ mJy/beam at 6.6 GHz, and scale with a factor of two. The horizontal bar indicates a length of 100 kpc at the cluster distance. The SRT beam at 4.8 GHz is plotted in the bottom-right corner.}
\label{fig.radio_maps}
\end{figure*}

In left panel of Fig.\ref{fig.radio_maps} we show an X-ray image taken from the ROSAT All-Sky Survey (RASS; Voges et al. 1999) compared with the contours of the SRT map at 4.8 GHz. 
The diffuse radio halo emission is clearly visible at the cluster centre, delimited by the $3\sigma$ contours.  
We note that this is the first time the radio halo in this cluster is detected at this frequency at this confidence level.

In the right panel of Fig.\ref{fig.radio_maps} we show the SRT image of the region of the radio halo at 4.8 GHz, overlapped with the 4.8 GHz contours and with the contours at 6.6 GHz from Murgia et al. (2024).  
The radio halo at the two frequencies has a similar extension, with a largest linear size of the order of $\sim22$ arcmin, but covering slightly different sky regions; the extension in the north direction is evident in both the images.  
We note that differences in the morphology of this halo at different frequencies have already been detected 
(see for example fig.4 in Thierbach et al. 2003 with the comparison between the 1.4 GHz and the 2.675 GHz maps). These differences mainly concern regions with low surface brightness. 
In these regions, differences in the halo spectral index spatial distribution, or fluctuations due to residual RFI, when  
observed with instruments having different angular resolution and noise level, can bring different parts of the halo just above or below the detection threshold.

For measuring the flux density of the halo, we used the proprietary software Synage++ (Murgia 2001). We first calculated the average surface brightness inside the $3\sigma$ contour defining the halo border, obtaining a value of 16.15 mJy/beam; we also calculated that the area of this region corresponds to 16.38 beams. We therefore estimated the total flux density of the region  
by multiplying the average surface brightness by the number of beams. In order to estimate the error on the flux density, we added in quadrature  
the rms noise of the image $\sigma_{rms}$ multiplied by the square root of the number of beams $N_b$ and  
a factor of $10\%$ of the derived flux density $F$, which takes into account possible systematic errors:
\begin{equation}
\sigma_F=\sqrt{(\sigma_{rms}\sqrt{N_b})^2+(0.1F)^2} .
\end{equation}
The result we obtained, which includes the contribution of discrete sources, is $265\pm10$ mJy. 

\begin{figure}
\centering
\begin{tabular}{c}
\includegraphics[width=\columnwidth, trim={1.5cm 11cm 0cm 1cm}, clip]{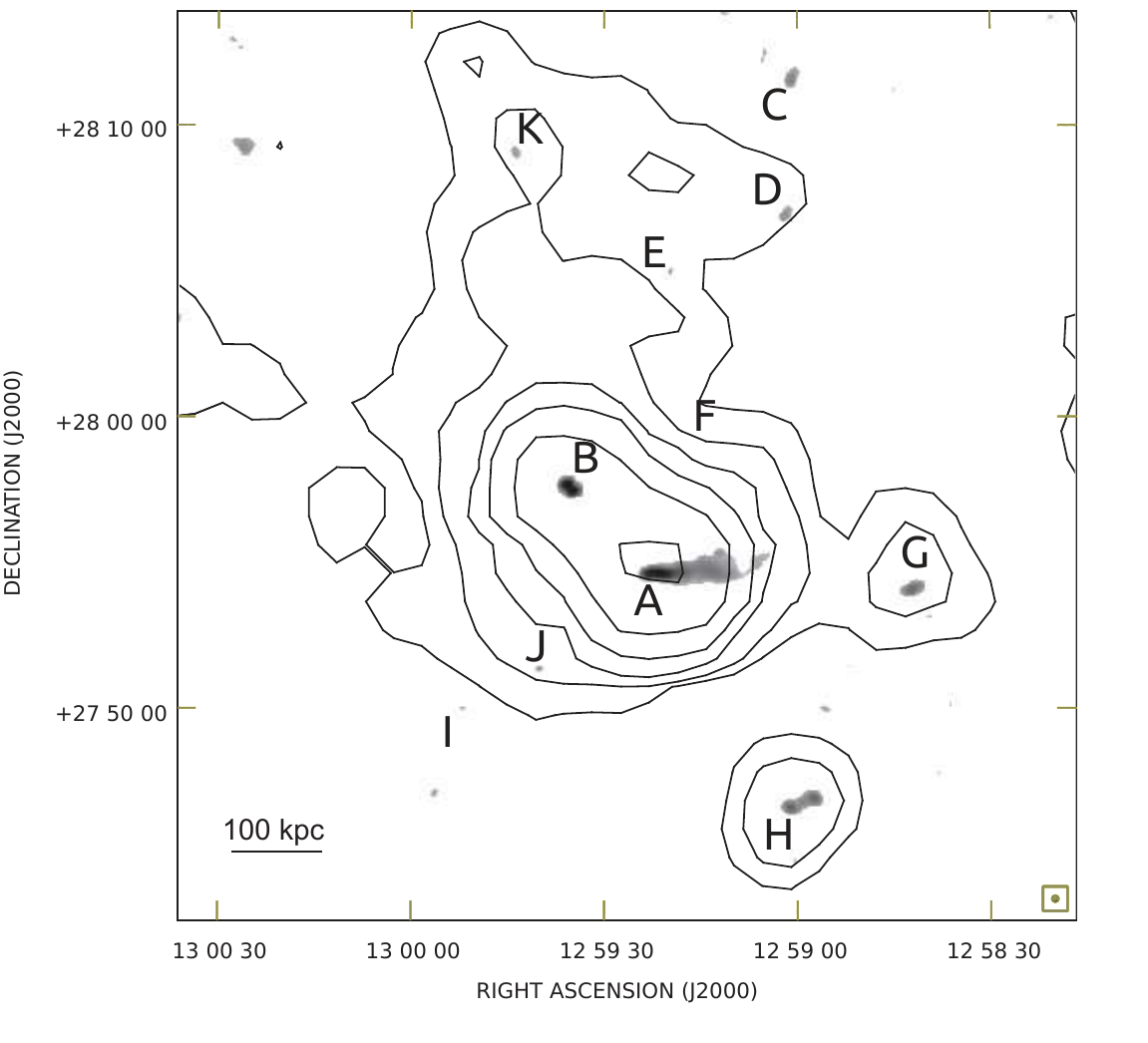}
\end{tabular}
\caption{VLA C-configuration image at 1.49 GHz with overlapped the 4.8 GHz SRT contours drawn as in right panel of Fig.\ref{fig.radio_maps}; sources labels are as in fig.5 in Murgia et al. (2024), with the addition of the K source (1257+28W06). The F source is not well visible in the image because it is covered by a contour line. The horizontal bar indicates a length of 100 kpc at the cluster distance. The VLA beam (13 arcsec) is shown in the bottom-right corner.}
\label{fig.vla_overlap}
\end{figure}

We adopted a procedure to subtract the contribution of discrete sources analogue to that of Murgia et al. (2024). We considered the sources in the list of that paper (see their table 2) located inside the $3\sigma$ contour of the halo region at 4.8 GHz (see Fig.\ref{fig.vla_overlap}, where the SRT contours at 4.8 GHz are overlapped on an archival VLA image in C-configuration at 1.49 GHz, from program AF196), and estimated the flux density of each source in two ways. For the sources for which a measure at $\sim4.8$ GHz is available in literature, we took that value, while for the sources where such a value is not available, we calculated it using the power-law fitting formulae reported in that paper (we note that for the two sources A and B, where a Continuous Injection model was necessary to fit the data, the measured flux densities at 4.8 GHz were available, so no estimations using that model have been necessary). As mentioned, we considered the sources located inside the $3\sigma$ contour of the halo at 4.8 GHz, so we excluded the sources labeled as C, H, and I in the list of Murgia et al. (2024), but added the source labeled as K in Fig.\ref{fig.vla_overlap}, i.e. the source 1257+28W06, which is outside the halo contour at 6.6 GHz but inside the halo at 4.8 GHz. For this source we extrapolated the flux density at 4.8 GHz by using the best fit formula derived by Kim (1994)  
(see Table \ref{tab.sources} for details about all sources). We obtained a total contribution of the discrete sources of $208.3\pm3.8$ mJy, which, once subtracted from the total emission of the region, provides a residual flux density due to the diffuse component of $56\pm11$ mJy.

\begin{table}{}
\caption{Discrete sources flux densities at 4.8 GHz, labeled as in Fig.\ref{fig.vla_overlap}, subtracted from the total flux density of the halo region. In the ``Instrument'' column, the instrument name is reported for direct measurements, while ``Fit'' is reported when an extrapolation using a fitting formula has been used. The reference for flux densities and fitting formulae is Murgia et al. (2024) for all sources, except for the K source, where the fitting formula is from Kim (1994)}
\begin{center}
\begin{tabular}{|*{3}{c|}}
\hline 
Source label & Flux density (mJy) & Instrument \\
\hline 
 A & $114\pm3$ & VLA \\
 B & $74\pm2$ & VLA  \\
 D & $4.9\pm1.0$ & Fit \\
 E & $2.4\pm0.3$ & Fit \\
 F & $0.86\pm0.05$ & VLA \\
 G & $10.4\pm0.8$ & Fit \\
 J & $0.54\pm0.05$ & VLA \\
 K & $1.2\pm0.4$ & Fit \\
\hline
 \end{tabular}
 \end{center} 
 \label{tab.sources}
 \end{table}

The last step is the correction for the SZE. As pointed out by En\ss lin (2002), this correction is necessary because at this frequency the decrement of the Cosmic Microwave Background (CMB) surface brightness due to the inverse Compton scattering with the hot gas electrons is a non-negligible fraction of the brightness of the halo. Therefore, when the baseline of the image is subtracted by estimating it considering regions outside the cluster, a CMB brightness stronger than the one present in the direction of the cluster can be subtracted, leading to an underestimation of the flux of the halo. As pointed out by Brunetti et al. (2013), for a correct estimate of the SZE correction it is important that the baseline of the image is estimated by using regions of the sky far enough from the cluster not to be affected by a residual SZE;  
otherwise, the resulting correction may overestimate the resulting flux density. Since the SZE in Coma appears to be strong up to a radius of the order of 45 arcmin from the center (Ade et al. 2013), we observed a region with side of more than 2 degrees, so that this problem should be avoided. 

In the range of frequencies considered in this paper, the SZE has a spectral shape proportional to $ \nu^2$, as proper to the Rayleigh-Jeans part of the blackbody spectrum, and is not affected by relativistic corrections appearing at higher frequencies (e.g. Colafrancesco, Marchegiani \& Palladino 2003), so the SZE correction in terms of surface brightness can be estimated as:
\begin{equation}
\left( \frac{\Delta I_{SZ}}{\text{mJy/beam}} \right ) = \left( \frac{1}{340}\right)  \left( \frac{\nu}{\text{GHz}}\right)^2 \left( \frac{\Delta T_{RJ}}{\text{mK}}\right) \left( \frac{\Omega_{\text{beam}}}{\text{arcmin}^2}\right)
\end{equation}
(e.g. Birkinshaw 1999). We calculate this expression by adopting $\nu=4.8$ GHz, $\Delta T_{RJ}=2 \langle y \rangle T_{CMB}$, with $\langle y \rangle=(4.3\pm1.0)\times10^{-5}$, as derived by Murgia et al. (2024) from the maps of Ade et al. (2013), $T_{CMB}=2725.5$ mK (Fixsen 2009), and $\Omega_{\text{beam}}=1.133 \times 3.9^2$ arcmin$^2$; we obtain the result $\Delta I_{SZ}=0.273 \pm 0.064$ mJy/beam. By multiplying this value by the number of the beams of the halo area, we obtain the correction in terms of flux density, which is $4.5\pm1.0$ mJy. Adding this result to the flux density of the diffuse component previously derived, we obtain that the flux density of the radio halo is estimated as $61\pm11$ mJy.

\section{Discussion}

The flux density at 4.85 GHz measured by Thierbach et al. (2003),  
once corrected by the SZE, was estimated as $42\pm12$ mJy by Murgia et al. (2024)\footnote{We note that this is the value resulting after that a material mistake in the Murgia et al. (2024) paper has been corrected (M. Murgia, private communication)}.
The result we obtained is therefore higher, even if marginally compatible, than the value present in literature. The reasons of this difference are not perfectly clear. As previously mentioned, the flux density in Thierbach et al. (2003) was obtained by integrating over a region with surface brightness smaller than $3\sigma$. It is not well clear from that paper the exact extension of the integration region; this is a possible source of difference, affecting also the amount of the SZE correction. Moreover, the subtraction of the discrete sources contribution was done using a method similar to the one of this paper, using as reference list the one of Kim (1994), but considering a higher number of sources. In fact, in that case  
the considered region was larger than ours, and more sources were present, obtaining an estimate of the contribution of discrete sources higher than ours, i.e. 230 mJy.  

The value we found for the flux density at 4.8 GHz can be compared with the one derived at 6.6 GHz by Murgia et al. (2024), which is $42\pm10$ mJy. As previously described, we followed a procedure to estimate the halo flux density as similar as possible to the one followed in that paper, by using the same methods for estimating and subtracting the fluxes of the discrete sources, and for correcting for the SZE. Therefore we believe it is appropriate to compare the two values, even if the different spatial distribution of the halo surface brightness at the two frequencies could bias the result; this is a general problem possibly affecting the whole spectrum (see, e.g., discussion in Brunetti et al. 2013). 

The integrated spectral index that can be estimated between the two frequencies is $\alpha_{4.8}^{6.6} = 1.17\pm0.94$.  
This value is
affected by a large relative error, because the two frequencies are quite close in a log scale, and therefore relatively small errors on the flux densities ($18\%$ and $24\%$ at 4.8 and 6.6 GHz respectively) produce a larger error on the spectral index ($\sim80\%$). Focusing on the central value of the spectral index, we note it is
 flatter than the one obtained by Deiss et al. (1997) by fitting the whole spectrum between 30 and 2700 MHz, $\alpha=1.36\pm0.03$ , and similar to the one obtained by the same authors by fitting the data until 1.4 GHz, $\alpha=1.16\pm0.03$. Considering the flux density at 1.49 GHz obtained by Murgia et al. (2024), $424\pm7$ mJy, 
we can estimate a spectral index of $\alpha_{1.49}^{4.8}=1.66\pm0.15$ between 1.49 and 4.8 GHz, suggesting that the steepening of the radio halo spectrum visible after 1.4 GHz seems to slow down  
after 4.8 GHz (see Table \ref{tab.spectr.ind} for a comparison between the values of the spectral index calculated between different frequencies), even if, given the large error on the spectral index between 4.8 and 6.6 GHz, this should be taken as a possible hint.

\begin{table}{}
\caption{Radio halo spectral index values derived between different couples of frequencies or from large band fitting; the note D97 indicates results of fits performed by Deiss et al. (1997), while the note M24 indicates spectral indices calculated using the flux densities derived by Murgia et al. (2024) at 1.49 and 6.6 GHz and the one obtained in this paper at 4.8 GHz}
\begin{center}
\begin{tabular}{|*{3}{c|}}
\hline 
Frequency band & Spectral index & Note \\
\hline 
 30--2700 MHz & $1.36\pm0.03$ & D97 \\
 30--1400 MHz & $1.16\pm0.03$ & D97 \\
 1.49--4.8 GHz & $1.66\pm0.15$ & M24 \\
 4.8--6.6 GHz & $1.17\pm0.94$ & M24 \\
\hline
 \end{tabular}
 \end{center} 
 \label{tab.spectr.ind}
 \end{table}

We note that a high-frequency spectral flattening is expected if the diffuse re-acceleration associated with the cluster turbulence has an important role in the origin of the radio halo, and if the seed electrons have a spectrum extending up to high energies,  
as in the case of continuous injection of secondary electrons by hadronic interactions or dark matter annihilation. In fact, since the re-acceleration characteristic time is approximately constant with the electrons energy $E_e$ (e.g. Brunetti et al. 2004), while the characteristic time of radiative energy losses is proportional to $E_e^{-1}$ (e.g. Sarazin 1999), there is an energy over which the energy losses become the dominant process even when the re-acceleration is active, and the electrons spectral shape recover that of the seed electrons, given by the equilibrium between continuous injection and energy losses. This reflects in the radio spectrum as a steepening until a certain frequency, and a following slowing down of the steepening if the seed electrons have an equilibrium spectrum with a relatively flat spectral index (see, e.g., Brunetti \& Lazarian 2011; see also Marchegiani et al. 2025 for a recent discussion in the case of electrons produced by dark matter annihilation).

Given the spectral index of the radio halo,
one can estimate the properties of the electrons and infer some conclusions about the processes that originated them. For example,  
we can consider the case where the electrons are produced in hadronic interactions between the nuclei of the ICM and non-thermal protons. If we assume that the non-thermal protons have a  
spectrum at the cluster center $N_p(\gamma)=N_{p,0}\gamma^{-s_p}$,
where $N_{p,0}$ is the normalization factor of the protons density at the cluster center and $s_p$ is their spectral index,
 the estimated radio spectral index along the whole spectrum, $\alpha\sim1.36$ (Deiss et al. 1997), 
would correspond to a spectral index of non-thermal protons of $s_p\sim2.7$ if the secondary electrons arise from the equilibrium between continuous production and energy losses (e.g. Marchegiani et al. 2007). 
In fact, in this case the spectral index of the electrons at equilibrium is given by $s_e=s_p+1$, and it is related to the synchrotron spectral index by $\alpha=(s_e-1)/2$ (e.g. Longair 1994), and therefore $s_p=2\alpha$. 

For this value of the protons spectral index, we calculated  
the radio spectrum produced by secondary electrons for a magnetic field with  
a central value of 4.7 $\mu$G and a radial profile $\propto n_{th}^{1/2}$ (Bonafede et al. 2010), where $n_{th}$ is the profile of the thermal gas derived from X-ray measures (Briel, Henry \& Boehringer 1992), and assuming that the non-thermal protons have the same radial profile of the thermal gas, extending up to a radius of $\sim0.5$ Mpc. 
This value, corresponding to $\sim17$ arcmin at the Coma distance, is higher than the halo radius found at high frequencies, but smaller than the one found at smaller frequencies (e.g. Bonafede et al. 2022). As outlined by, e.g., Brunetti et al. (2013), the measured size of the halo is different at different frequencies when using instruments with different sensitivity, angular resolution, and angular scales coverage, and this is a possible explanation for the dispersion of the spectral data. The value used here is therefore a simplifying approximation.

\begin{figure*}
\centering
\begin{tabular}{c}
\includegraphics[width=0.5\textwidth, trim={2.2cm 13cm 0.8cm 4.2cm}, clip]{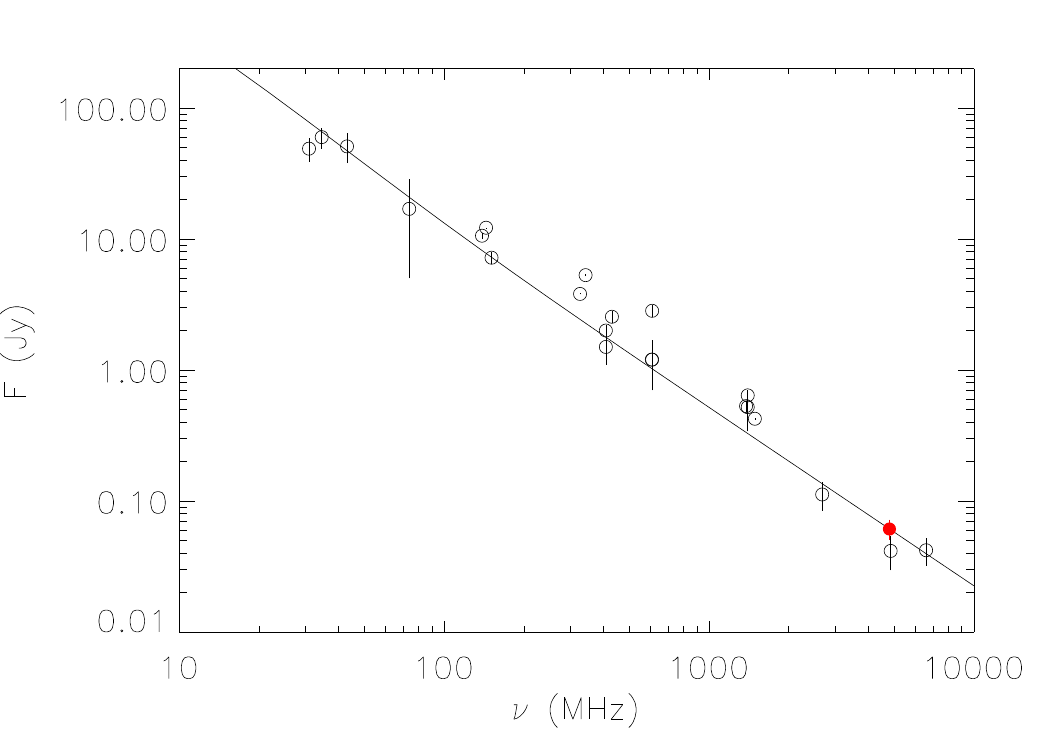}
\includegraphics[width=0.5\textwidth, trim={2.2cm 13cm 0.8cm 4.2cm}, clip]{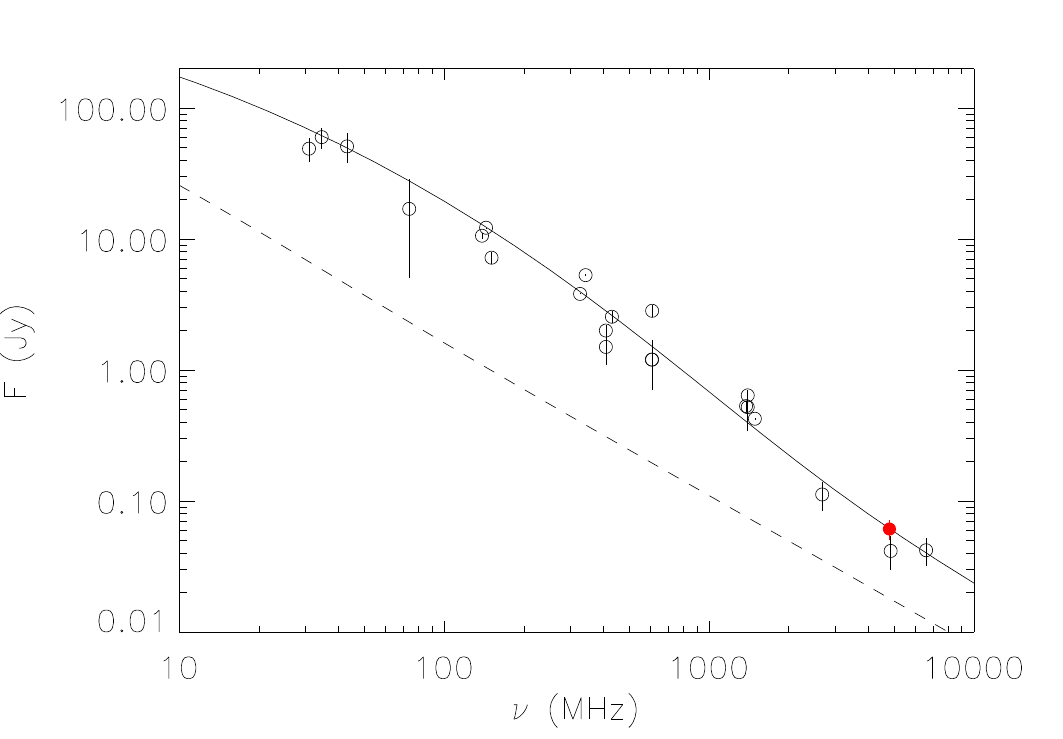}
\end{tabular}
\caption{\textit{Left panel:} Spectrum of the Coma radio halo produced by secondary electrons of hadronic origin for a spectral index of non-thermal protons $s_p=2.7$ and a central density normalization of $N_{p,0}=5.5\times10^{-9}$ cm$^{-3}$, in absence of turbulent re-acceleration. \textit{Right panel:} As in the left panel, but for $s_p=2.3$, $N_{p,0}=1.5\times10^{-10}$ cm$^{-3}$, and for re-acceleration with $\chi=5\times10^{-17}$ s$^{-1}$ and $T_{acc}=3\times10^8$ yr (solid line) and without re-acceleration (dashed line). Empty circles are data taken from the compilation in Murgia et al. (2024), while the filled red circle is from this paper.}
\label{fig.spectra}
\end{figure*}

With these assumptions, we can constrain the central density of the non-thermal protons by adjusting the value of $N_{p,0}$. In the left panel of Fig.\ref{fig.spectra} we show that  
a normalization of the  non-thermal protons spectrum of $N_{p,0}=5.5\times10^{-9}$ cm$^{-3}$ can produce a radio spectrum similar to the observed one in absence of re-acceleration.  
However, the gamma ray flux above 100 MeV produced in hadronic interactions,
calculated as described in Marchegiani et al. (2007) and references therein, 
 in this case results to be $F(>100 \mbox{ MeV})=3.9\times10^{-9}$ cm$^{-2}$ s$^{-1}$, which is very close 
to the \textit{Fermi}-Large Area Telescope (LAT) upper limit of $F_{UL}=4.2\times10^{-9}$ cm$^{-2}$ s$^{-1}$ (Ackermann et al. 2016). This model is therefore problematic.

If we consider that secondary electrons can be the seed for turbulent re-acceleration (see, e.g., Brunetti et al. 2012), we can assume that the spectrum between 4.8 and 6.6 GHz reflects the one due to the seed electrons, and calculate the spectrum produced in the presence of re-acceleration. Therefore, 
if we use  
the central value of the spectral index we have found, $\alpha_{4.8}^{6.6} \sim 1.17$, we can assume a spectral index for non-thermal protons of $s_p=2.3$.
In this way, it is possible to obtain a better explanation of the data by assuming a smaller density for the protons.  
In fact, in the right panel of Fig.\ref{fig.spectra} we show the spectrum obtained for $N_{p,0}=1.5\times10^{-10}$ cm$^{-3}$, if a turbulent re-acceleration with acceleration parameter $\chi=5\times10^{-17}$ s$^{-1}$ and duration $T_{acc}=3\times10^8$ yr (see Marchegiani et al. 2025 for details about the meaning of these quantities) is assumed. In the same plot, we also show for reference the spectrum produced in absence of re-acceleration for the same values of protons density and spectral index. In this case, the estimated gamma ray flux is $F(>100 \mbox{ MeV})=2.1\times10^{-10}$ cm$^{-2}$ s$^{-1}$, much lower than the \textit{Fermi}-LAT upper limit. 

It is important to note that the models presented in Fig.\ref{fig.spectra} are not obtained through a fitting procedure to the data, but, once that some parameters (protons spectral index, spatial distribution, magnetic field) are assumed, just by adjusting by hand the value of $N_{p,0}$ and, for the case with re-acceleration, the values of $\chi$ and $T_{acc}$. In fact, the data have a large dispersion, due to the different sensitivities and angular resolution of the different instruments, and the different techniques for estimating the flux density and the corresponding error bar adopted in different papers. Therefore using a fitting technique based, for example, on the minimization of the $\chi^2$ parameter, would not produce acceptable results (see, e.g., discussions in Brunetti et al. 2013, Murgia et al. 2024, Marchegiani et al. 2025). The different models are here shown for a comparison by eye between theoretical models and data, without a quantitative evaluation of their goodness.
Moreover, we outline that these results are obtained under the assumption that the spectrum of the non-thermal protons is not changed by the effect of the turbulent re-acceleration, which is an untrue hypothesis (e.g. Brunetti et al. 2012), so these results should be taken just as a first rough estimate of the parameter space we should explore to obtain a consistent explanation of the data, which can be done in future works.

\section{A polarized spot inside the radio halo}

As mentioned in Sect.1, Murgia et al. (2024) reported at 6.6 GHz the presence inside the halo of a polarized spot, without a clear counterpart in the total intensity map at the same frequency, nor in maps at lower frequencies. The fractional polarization $FPOL=P/I$ of this spot was around 45\% for an angular resolution of 2.9 arcmin.

\begin{figure}
\centering
\begin{tabular}{c}
\includegraphics[width=\columnwidth, trim={1.5cm 9.5cm 0cm 0cm}, clip]{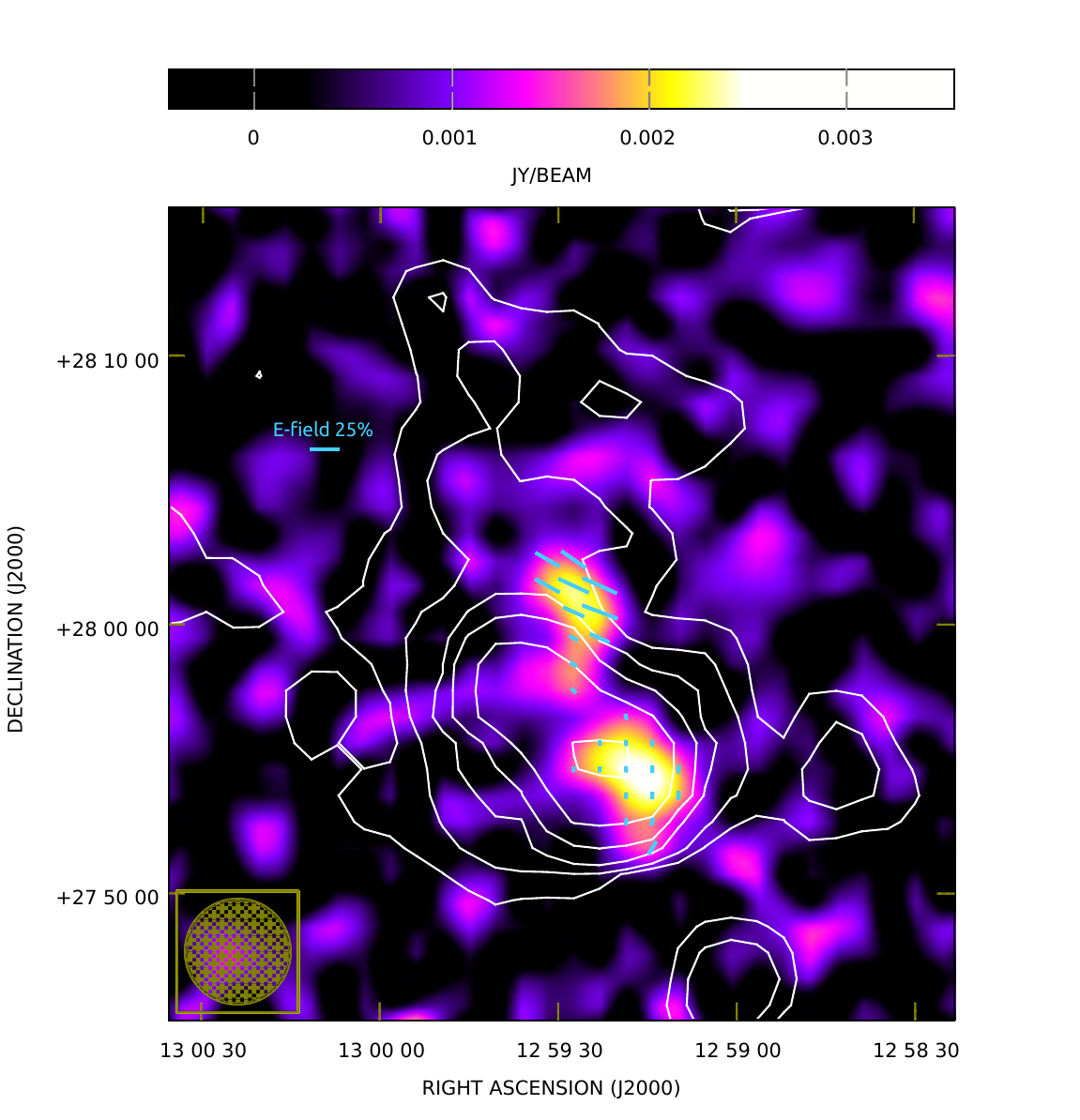}
\end{tabular}
\caption{Linearly polarized intensity map, interpolated with the Laplace-Everett method, at 4.8 GHz obtained with SRT of the halo region in Coma; overlapped are the contours of the total intensity map, starting at $3\sigma_I$, with $\sigma_I=1$ mJy/beam, and scaling with a factor of two. The polarized map is cut at $3\sigma_P$, with $\sigma_P=0.5$ mJy/beam. The vectors refer to the radio wave E-field and are traced only where the fractional polarization $FPOL=P/I$ is detected with signal/noise higher than 3. The SRT beam at 4.8 GHz is plotted in the bottom-left corner.}
\label{fig.radio_pol}
\end{figure}

While previous Sections were focused on the results in total intensity, we present here
the map in polarized intensity, derived as described in Sect.2.  
We obtained an interesting result, i.e. that the polarized spot is also present at 4.8 GHz, as visible in Fig.\ref{fig.radio_pol}, where we show the polarized intensity map with overlapped the total intensity contours and the E-field vectors.

In the polarized image, there are two clearly visible features; the southern one corresponds to the radio galaxy NGC\,4869, while the other one does not seem to have an evident counterpart. This spot has an higher fractional polarization compared to the radio galaxy ($\sim25\%$ vs. $\sim3\%$), and the E-field vectors appear to be oriented at position angle $65\pm 5$ deg.

\begin{figure}
\centering
\begin{tabular}{c}
\includegraphics[width=\columnwidth, trim={1cm 6.5cm 2cm 6.5cm}, clip]{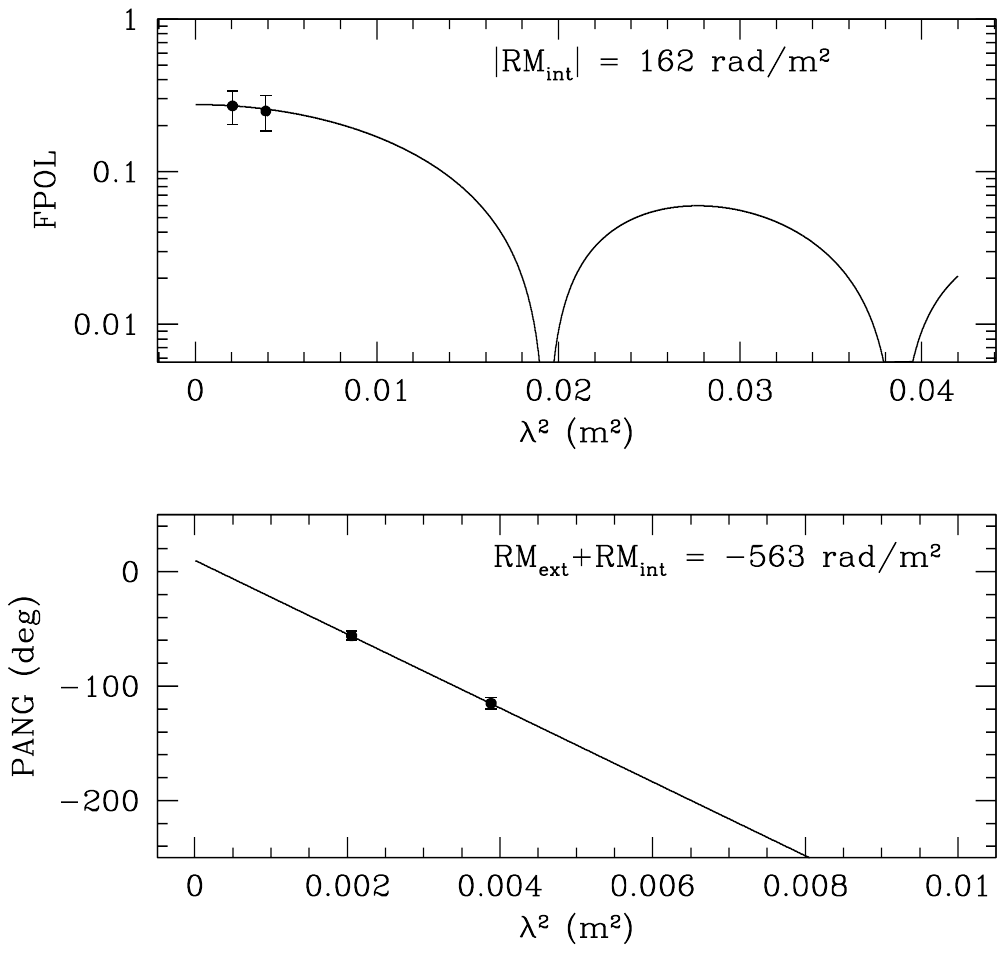}
\end{tabular}
\caption{\textit{Upper panel:} fractional polarization of the polarized spot as a function of the wavelength, fitted with an internal depolarization model. \textit{Lower panel:} polarization angle of the polarized spot as a function of the wavelength, fitted with a rotation measure model.}
\label{fig.rm}
\end{figure}

The presence of this feature is interesting, because can be in accordance with the simulations of Govoni et al. (2013), who found that the radio halos can be intrinsically polarized, but can appear generally unpolarized because of the effect of the resolution and the sensitivity of the instrument, with localized spots that can appear in regions where the magnetic field has a preferential orientation.  
The fractional polarization is expected to increase with the frequency; this trend is in line with the value found at 6.6 GHz by Murgia et al. (2024) and the value we found at 4.8 GHz. For a proper comparison, we convolved the 6.6 GHz image with the SRT 4.8 GHz beam, 3.9 arcmin, finding $FPOL_{6.6}\sim0.27$. 
For this value, the depolarization is $DP=FPOL_{4.8}/FPOL_{6.6}=0.25/0.27\sim0.93$.
This change in polarization fraction is compatible with internal polarization with $|RM_{int}|\sim160$\,rad/m$^2$ (see upper panel of Fig.\ref{fig.rm}), which can be provided, for example, by an uniform slab with magnetic field $B \sim2\,\mu$G and physical size $L=100$\,kpc.

By comparing the polarization angles measured at the two frequencies, we find $\Psi_{4.8}=65\pm5$\,deg and $\Psi_{6.6}=-56\pm4$\,deg. We can therefore estimate the quantity $RM=\Delta \Psi / \Delta \lambda^2$, which, taking into account the $n\pi$ degeneracy of the polarization angle measures, 
can be explained with $RM\sim-560$\,rad/m$^2$ (see lower panel of Fig.\ref{fig.rm}). 
By subtracting the internal $RM_{int}$ estimated above we obtain, choosing the sign of this last quantity to get the minimum possible value in modulus, $RM_{ext}\sim-400$\,rad/m$^2$, which is in line with the estimates in the cluster center region of Bonafede et al. (2010) (see, e.g., their fig.10).

This polarized spot therefore deserves a more in-depth study, using higher resolution images to better check the presence of a possible counterpart, in order to confirm, or exclude, that it is intrinsically part of the halo.

\section{Conclusions}

In this paper we have presented the results of the observations of the Coma radio halo at 4.8 GHz with the SRT; for the first time, the radio halo has been detected with a significance $>3\sigma$ at this frequency.

The derived flux density, after cleaning the map from RFI contamination, after removing the contribution of discrete sources, and after correcting for the SZE, is $61\pm11$ mJy. This number is higher than the flux density at the same frequency reported in literature (Thierbach et al. 2003), but marginally compatible when also this last result is corrected for the SZE.

The integrated spectral index between 4.8 and 6.6 GHz, obtained by comparing our result with the one by Murgia et al. (2024), is $\alpha\sim1.17$, suggesting that the spectrum of the radio halo, after the quick steepening observed after 1.4 GHz, is slowing down its steepening after 4.8 GHz. This behavior is compatible with scenarios where re-acceleration due to the cluster turbulence acts on seed electrons that are continuously produced, as secondary electrons of hadronic origin or produced by dark matter annihilation. Scenarios where the seed electrons have a relatively flat spectrum and are re-accelerated for a fraction of Gyr appears to be favored compared with models with steeper spectra and without re-acceleration, because they provide a better accordance with the radio spectral shape, and are less subject to violate the gamma ray upper limits.

We have also seen that the polarized spot inside the halo already detected at 6.6 GHz by Murgia et al. (2024) is present also at 4.8 GHz, and have derived some possible constraints on the properties that the magnetic field should have by looking at the differences in fractional polarization and polarization angle at the two frequencies.

\section*{Acknowledgments}

The Enhancement of the Sardinia Radio Telescope (SRT) for the study of the Universe at high radio frequencies is financially supported by
the National Operative Program (Programma Operativo Nazionale - PON) of the Italian Ministry of University and Research ``Research and
Innovation 2014-2020'', Notice D.D. 424 of 28/02/2018 for the granting of funding aimed at strengthening research infrastructures, in
implementation of the Action II.1 -- Project Proposals PIR01$\_$00010 and CIR01$\_$00010. 
We thank the referee for comments and suggestions.

\section*{Data availability}
The data underlying this article will be shared on reasonable request to the corresponding author.

\bsp

\label{lastpage}

\end{document}